\documentclass[10pt,letterpaper]{article}

\usepackage{graphicx}
\usepackage{epstopdf}
\usepackage[ansinew]{inputenc}
\usepackage{float}
\usepackage{textcomp}
\usepackage{subfigure}
\usepackage{color}
\usepackage{amsmath}

\usepackage{opex3}

\newcommand{\DW}[0]{{\mathrm{DW}}}
\newcommand{\Real}[0]{\mathrm{Re}}
\newcommand{\Imax}[0]{I_\mathrm{max}}
\newcommand{\Imin}[0]{I_\mathrm{min}}

\begin{document}

\title{Broadband, stable and highly coherent supercontinuum generation at telecommunication wavelengths in an hydrogenated amorphous silicon waveguide}

\author{F. Leo$^{1,2}$, J. Safioui$^3$, B. Kuyken$^{1,2}$, G. Roelkens$^{1,2}$ and  S.-P. Gorza$^3$}

\address{$^1$ Photonics Research Group, Department of Information Technology, Ghent University-IMEC, Ghent B-9000, Belgium\\
$^2$Center for nano- and biophotonics (NB-photonics), Ghent University, Belgium\\
$^3$ Service OPERA-photonique, Université libre de Bruxelles (ULB), 50
Avenue F. D. Roosevelt, CP194/5 B-1050 Bruxelles, Belgium.}

\email{sgorza@ulb.ac.be}

\begin{abstract}

Hydrogenated amorphous silicon (a:Si-H) has recently been
recognized as a highly nonlinear CMOS compatible photonic
platform. We experimentally demonstrate the generation of a
supercontinuum (SC) spanning over 500\,nm in a-Si:H photonic wire
waveguide at telecommunication wavelengths using femtosecond input
pulse with energy lower than 5\,pJ. Numerical modeling of pulse
propagation in the waveguide, based on the experimentally
characterized dispersion profile, shows that the supercontinuum is
the result of soliton fission and dispersive wave generation. It
is demonstrated that the SC is highly coherent and that the
waveguides do not suffer from material degradation under
femtosecond pulse illumination. Finally, a direct comparison of SC
generation in c-Si and a-Si:H waveguides confirms the higher
performances of a-Si:H over c-Si for broadband low power SC
generation at telecommunication wavelengths.
\end{abstract}



\section{Introduction}

Since its first discovery in the early seventies by Alfano and
Shapiro\,\cite{Alphano-70}, the phenomenon of supercontinuum
generation (SCG) has been thoroughly studied, particularly since
the advent of photonic crystal fibers (PCFs). This fascinating
nonlinear process gives rise in the extreme case to
octave-spanning spectra and found numerous applications in
frequency metrology, spectroscopy, optical communications, or
medical imaging\,\cite{Dudley-06}. PCF-based supercontinua are
nowadays routinely generated in laboratories and SC-sources are
commercially available. Recently, the generation of SC in
integrated photonics has attracted a lot of attention for its
potential for high-volume, low-cost and low power consumption,
integrated broadband or few-cycle pulse sources. On-chip
supercontinuum generation at wavelengths in the C band of
telecommunications has been reported in
chalcogenide\,\cite{Lamont-08}, silicon nitride\,\cite{Halir-12},
amorphous silicon\,\cite{Safioui-14} and silicon photonic
wires\,\cite{Hsieh-07, Ding-10, Leo-14}. The generation of broad
SC spectra in silicon chips is challenging near 1550\,nm because
two-photon absorption and subsequent free-carriers effects limit
the spectral broadening\,\cite{Leo-14}. To overcome this problem,
it has been shown that resorting to short pulses
($<100$\,fs)\,\cite{Ishizawa-14} or nanoscale slot
waveguides\,\cite{Zhang-12} improve the SC bandwidth. An
alternative is to seed the SC in the mid-IR where lower nonlinear
losses are experienced because the pump photon energy is below the
mid bandgap of crystalline silicon\,\cite{Kuyken-11a}. Note that
an octave-spanning SC, seeded in the mid-IR, has recently been
reported in a-Si:H\,\cite{Shen-14}.

Few years ago, hydrogenated amorphous silicon (a-Si:H) has been
identified as a very promising material for nonlinear optics at
telecommunication wavelengths, either in photonic
chips\,\cite{Ikeda-07, Narayanan-10, Shoji-10, Kuyken-11, Wang-12}
or in optical fibers\,\cite{Mehta-12}. As crystalline silicon
(c-Si) waveguides, a-Si:H photonic wires can be fabricated with
CMOS compatible processes on silicon-on-insulator. However it has
the advantage over c-Si of being characterized by a higher
nonlinear refractive index and a lower nonlinear absorption
coefficient resulting from a larger bandgap energy. Thanks to
these favorable properties, the generation of a broad
supercontinuum, seeded by picosecond pulses, in a-Si:H waveguides
with low anomalous dispersion at 1550\,nm has been
demonstrated\,\cite{Safioui-14}. In this latter work, the initial
spectral broadening is the result of spontaneous generation of new
frequencies from noise. The generated SC are thus expected to be
incoherent with large amplitude and phase fluctuations from
shot-to-shot. In this work, we study the properties of
supercontinua seeded by shorter pulses ($<200$\,fs). In this
regime, the initial spectral broadening is expected to be
dominated by self-phase modulation\,\cite{Dudley-06} and therefore
a highly coherent SC should be generated.

\section{Supercontinuum generation}

\begin{figure}[] 
\vspace{-0mm}
\centerline{\resizebox{100mm}{!}{\includegraphics{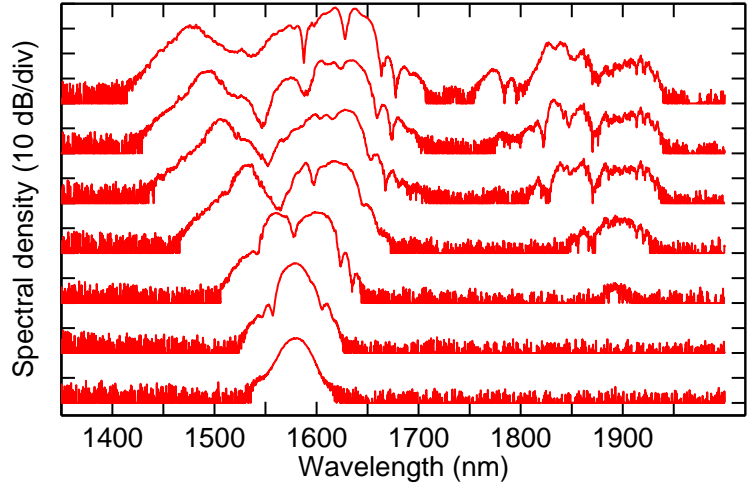}}}
\caption{Spectra recorded at the output of a 1-cm-long a:Si-H
waveguide for an input on-chip peak power of 0.03\,W, 1.2\,W,
2.3\,W, 4.6\,W, 6.9\,W, 9.8\,W and 13\,W (bottom to top). The
curves are shifted by 20 dB for clarity.} \label{Fig_1}
\vspace{-2mm}
\end{figure}

In our experiments, we consider 1-cm-long, 220-nm-thick and
500-nm-wide hydrogenated amorphous silicon (a:Si-H) photonic wire
waveguides. The a-Si:H film was deposited using a low temperature
Plasma Enhanced Chemical Vapor Deposition (PECVD) process on top
of a 1950 nm thick layer of high-density plasma oxide on a silicon
substrate\,\cite{Kuyken-11}. The input pump was the idler output
of an OPO laser pumped by a Ti:Sapphire laser (Spectra Physics
OPAL and Tsunami, respectively), which delivers 180\,fs pulses at
82\,MHz repetition rate. The horizontally-polarized laser output
was coupled into the waveguide by means of a $\times60$ microscope
objective (NA = 0.65) to excite the quasi-TE mode of the
waveguide. At the output, the light was collected by a lensed
fiber (NA = 0.4) and sent either into an optical spectrum analyzer
or to an interferometer to measure the coherence of the generated
SC. The input and output coupling efficiencies were measured to be
-20.8\,dB and -9.5\,dB respectively.

Typical SC spectra at the waveguide output are reported in
Fig.\ref{Fig_1} for seed pulses at 1575\,nm and for an on-chip
input power ranging from 0.03 to 13\,W. As can be seen, as the
input power increases, the pump spectrum broadens by self-phase
modulation and, at input powers larger than 2.3\,W, the spectrum
exhibits features far from the pump wavelength around 1900\,nm. At
high power, i.e. for a pump energy close to 3\,pJ, the spectrum
extends from 1450 to 1850\,nm at -20 dB. Note that
in\,\cite{Gorza-14} we have reported larger SC spectra spanning
from 1380 nm to 1920 nm at -20 dB, at slightly higher power
(18\,W) and in a similar but not identical waveguide. Referring
to\,\cite{Dudley-06}, at high power, the SC spectrum is likely the
result of self-phase modulation, soliton fission and dispersive
wave generation. The measurement of the output power as a function
of the input power (not shown) reveals that the nonlinear
absorption is not negligible and saturates the available output
energy in the supercontinuum.

\section{Group velocity dispersion measurement}\label{Section2}

\begin{figure}[] 
\vspace{-0mm}
\centerline{{\includegraphics[width=3.8cm]{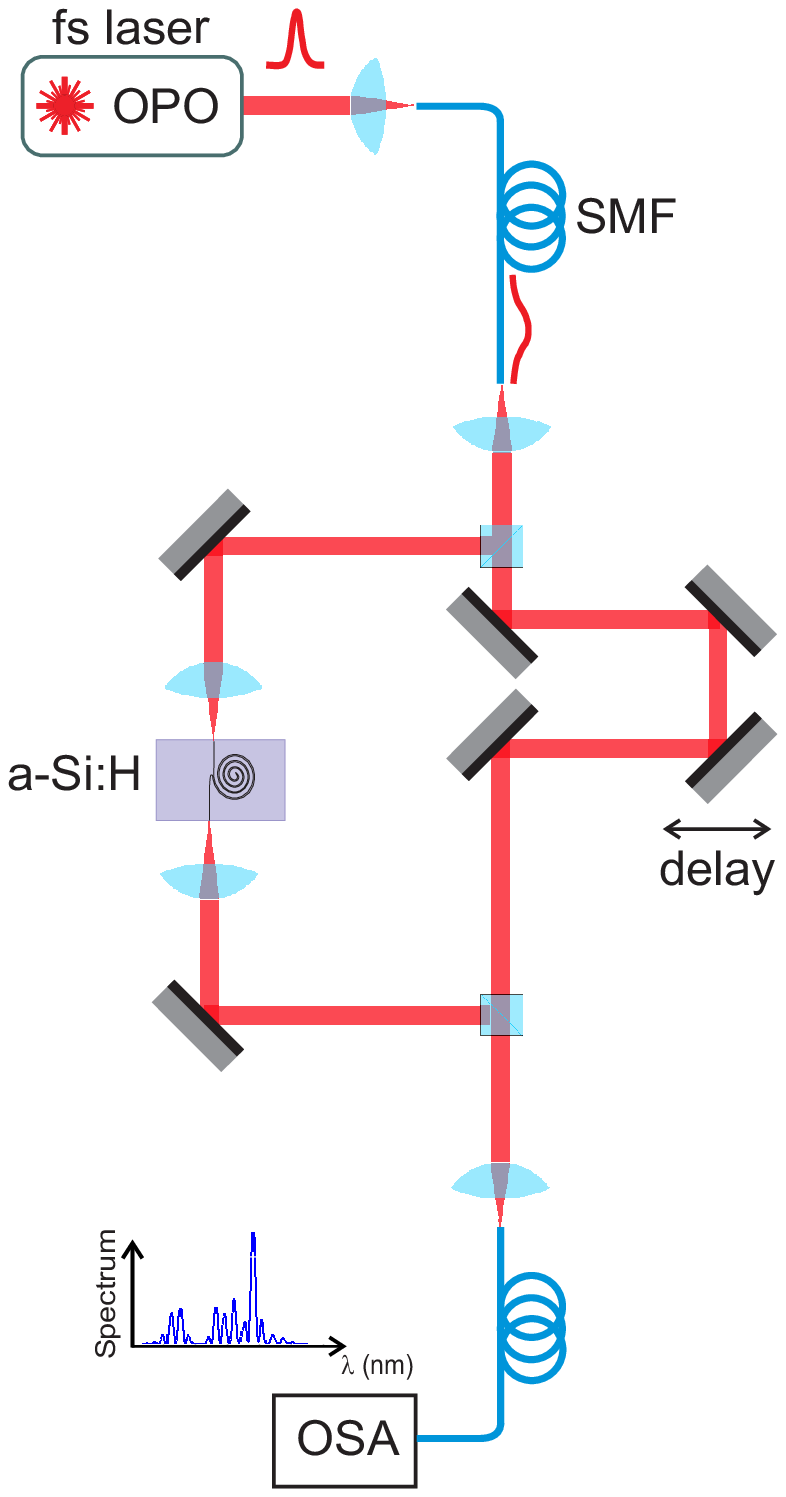}\hfill
\includegraphics[width=9cm]{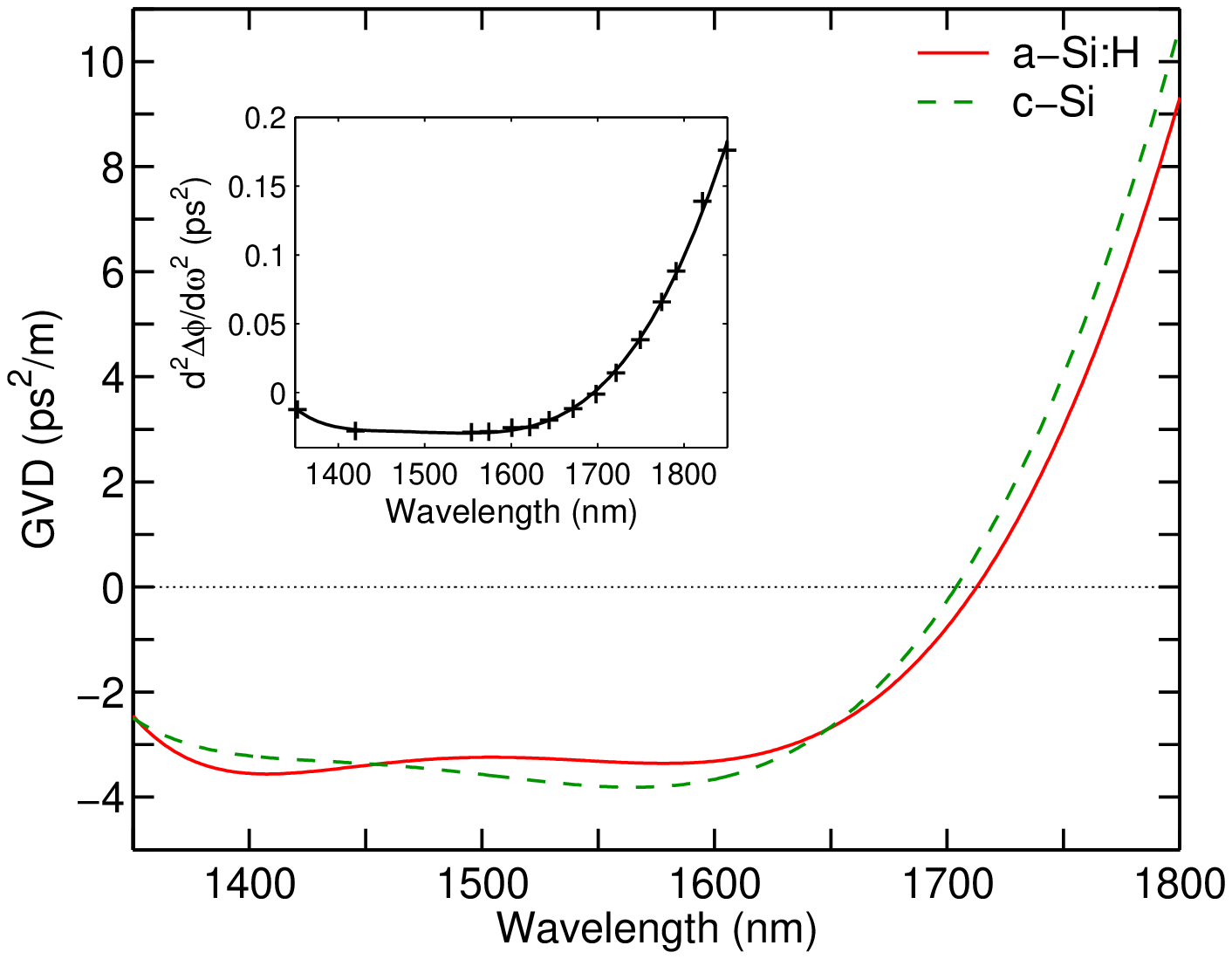} }} \caption{Left:
Experimental setup used to measure the GVD of our nanowires.
Right: Measured GVD. The solid red curve shows the fourth-order
polynomial fit of $\beta_2(\omega)$ plotted as a function of the
wavelength, for our a:Si-H waveguides. The curvature of the phase
difference between the reference and the waveguide arms
($d^2\Delta\phi /d\omega^2$) for the 1-cm-long photonic wire is
shown in the inset (cross) along with the 4th order polynomial fit
of the data (solid line). The green dashed line shows the measured
GVD of similar crystalline silicon waveguides (same mask).}
\label{Fig_2} \vspace{-2mm}
\end{figure}

The dispersion properties of the waveguide play an important role
in the SC generation. The spectral characteristics of the SC shown
in Fig.\ref{Fig_1} can thus be fully understood and reproduced in
numerical simulations only if the group velocity dispersion (GVD)
is known over its whole spectrum. a-Si:H is a material with
considerable variation in chemical structure, and different
fabrication processes might lead to different material and hence
optical properties. In the absence of sufficient data about the
refractive index of our a-Si:H film in the wavelength range of
interest, we have measured the dispersion properties of the
waveguide. The GVD was measured at different wavelengths by
resorting to standard spectral interferometric
methods\,\cite{Merritt-89,Leo-14c}. We used an unbalanced free
space Mach-Zehnder interferometer in which, in one arm, the light
propagates in the waveguide under test (see Fig.\,\ref{Fig_2}).
The light source was the signal ($\sim 1350$\,nm) or the idler
($1550-1750$\,nm) output of the OPO laser, or alternatively a SLED
light source (1500-1600\,nm). In order to avoid any nonlinear
effects in the waveguide, the pulses at the output of the OPO were
temporally broadened by propagation through SMF fiber. At the
output of the interferometer, the spectral fringes between the
pulses from the two arms were recorded with an optical spectrum
analyzer. Thanks to the nonzero delay between the two pulses,
their spectral phase difference ($\Delta{\phi}(\omega)$) can
readily be extracted by numerically filtering the Fourier
transform of the interference pattern\,\cite{Iaconis-99}. The
inset in Fig.\,\ref{Fig_2} shows the second order derivative
$d^2\Delta\phi /d\omega^2$ as a function of the wavelength, for
the 1-cm-long waveguide. Each cross corresponds to one
measurement. This data were fitted by a fourth order polynomial
curve in the frequency domain. In order to extract the GVD, the
measurements were done for 1 and 2 cm-long waveguides. The GVD
plotted in Fig.\,\ref{Fig_2} is thus given by
$\beta_2(\omega)=[d^2\Delta{\phi}_{2cm}/d\omega^2-d^2\Delta{\phi}_{1cm}/d\omega^2]/1\mathrm{cm}$.

It can be seen that at telecommunication wavelengths, the
dispersion is anomalous and is about -3.3\,ps$^2$/m. This value is
in agreement with previous results based on the estimation of
$\beta_2$ from a four-wave-mixing experiment\,\cite{Kuyken-11} in
similar waveguides. The zero-dispersion wavelength is measured to
be about 1710\,nm, meaning that part of the SC spectrum reported
in Fig.\ref{Fig_1} is in the normal dispersion regime. This
suggests that the spectral broadening is partially the result of
the resonant energy transfer from the strong pump pulse
propagating in the anomalous GVD regime to a \v{C}erenkov
dispersive wave in the normal dispersion
regime\cite{Akhmediev-95}. The phase-matching condition associated
with dispersive wave emission  is $\sum_{k \ge
2}\beta_k(\omega_{\DW}-\omega_s)^k/k!=\gamma P$, where $\beta_k$
are the dispersion coefficients associated with the Taylor series
expansion of the propagation constant around the pump frequency
$\omega_s$ and $\omega_{\DW}$ is the frequency of the dispersive
wave. This condition predicts the dispersive wave to be emitted at
$\lambda_{\DW}=1950$\,nm in good agreement with the small spectral
peak seen in Fig.\,\ref{Fig_1} at 2.3\,W peak power. This also
confirms the results reported in\,\cite{Safioui-14}, where SC
generation was investigated in the picosecond regime in similar
waveguides.

\section{Numerical simulations}

Further insights about the underlying physics of SC generation in
the a-Si:H waveguides displayed in Fig.\,\ref{Fig_1} can be gained
by numerically simulating the propagation of the optical field.
The model considered is the generalized nonlinear Schrödinger
equation (GNLSE), where higher order dispersion and
self-steepening terms are included:
\begin{equation}
\frac{\partial A(z,t)}{\partial z}= i\sum_{k=2}^6  \frac{i^k}{k!}\beta_k\frac{\partial^k A(z,t)}{\partial t^k} -\frac{\alpha_0}{2}A(z,t)+i(1+\frac{i}{\omega_0}\frac{\partial}{\partial t})A(z,t)\int_{-\infty}^{\infty}R(t')\vert A(z,t-t')\vert^2 dt'.\\
\label{Eq1}
\end{equation}
In this equation $A(z,t)$ is the slow-varying envelope of the
electric field, $R(t) = \gamma \delta(t)$ with
$\gamma=(770+i30)$\,W$^{-1}$m$^{-1}$\,\cite{Kuyken-11} and the
coefficients $\beta_k$ are extracted from the measurements
reported in Fig.\,\ref{Fig_2}. The linear losses were measured to
be 2.65\,dB/cm by cutback method.

Raman scattering has been reported both in a-Si and in
a-Si:H\,\cite{Maley-88, Mehta-10}. However, its weakness prevented
us to see evidence of the Raman contribution in previous studies
with our a-Si:H photonic wires. The Raman effect was thus
neglected in Eq.\ref{Eq1}. In crystalline silicon, where
free-carrier index change (FCI) and free-carrier absorption (FCA)
are well known, it has been shown that FCI and FCA do not
sgnificantly modify the SC generation in the femtosecond
regime\,\cite{Leo-14}. We have thus not considered the FCA and FCI
contribution. Finally note that it has recently been reported in
\cite{Wathen-14} that hydrogenated amorphous silicon waveguides
might show absorptive and refractive non-instantaneous
nonlinearity\,\cite{Ikeda-07} as well as no instantaneous
nonlinear absorption. We have not included these results in our
model because these effects are not known for our waveguide and
might strongly depend on the fabrication process as suggested by
the variety of nonlinear characteristics of a-Si:H waveguides
reported in literature (see e.g.\,\cite{Shoji-10}).

\begin{figure}[]  
\vspace{-0mm}
\centerline{\subfigure[P=2.3W]{\includegraphics[width=7cm]{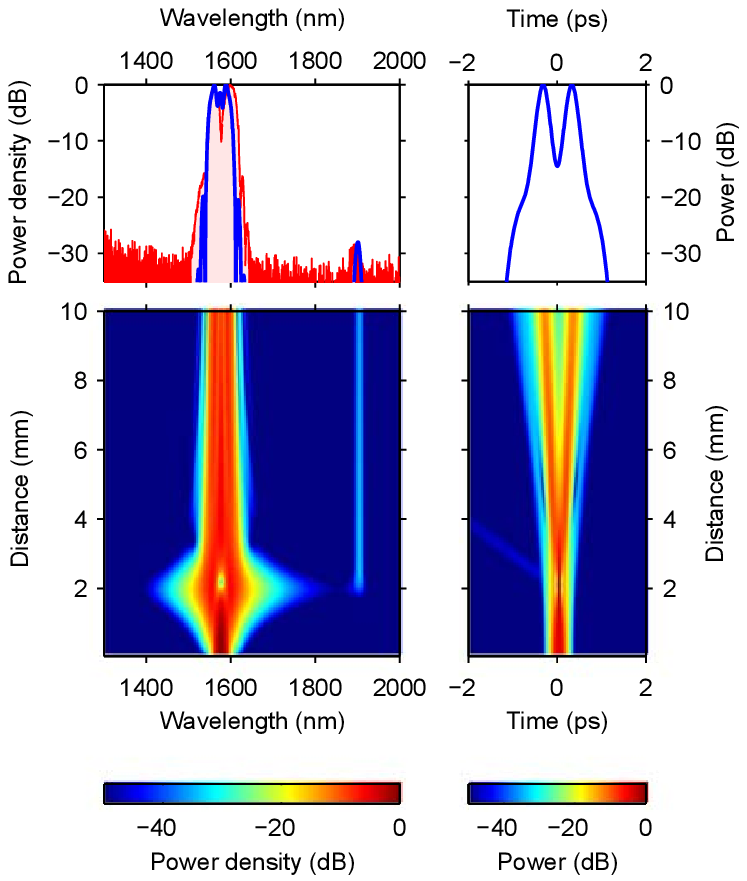}}\hfill
\subfigure[P=13W]{\includegraphics[width=7cm]{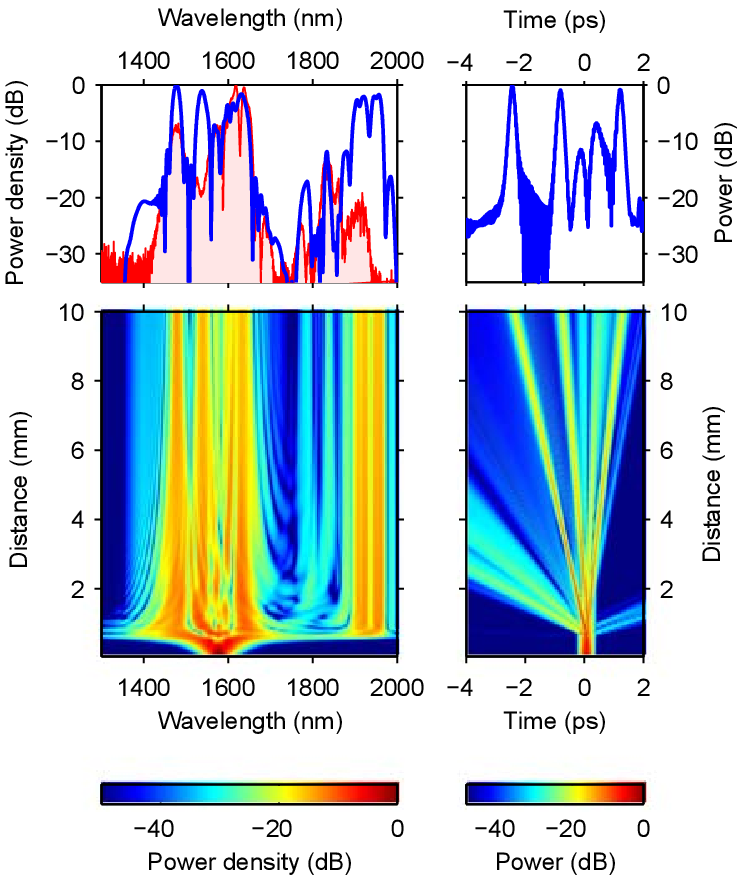}
}} \caption{Pseudocolor plots of the simulated spectral and
temporal evolution along the propagation for a 2.3\,W (left) and
13\,W (right) 188\,fs Gaussian input pulse (FWHM). Top plots
(blue) highlight the waveguide output at z = 10 mm. The red filled
curve is the measured output spectrum for comparison (see also in
Fig.\,\ref{Fig_1}).} \label{Fig_3} \vspace{-2mm}
\end{figure}

The measurement of the autocorrelation trace and the spectrum of
the input pulses agree well with a Gaussian pulse shape
$E=A_0exp(-[t/T_0]^2)$ with $T_0=160$\,fs. At a moderate peak
power of 2.3\,W, i.e. for a soliton number
$N=[T_0^2\Real\{\gamma\}P_0/(2\beta_2)]^{1/2}=2.6$, the numerical
simulation shows that the pump spectrum is broaden by self-phase
modulation in the first millimeters of propagation (see
Fig.\,\ref{Fig_3}(a)). This broadening is accompanied by a
temporal compression and is followed by a splitting of the initial
high-order soliton in two solitons with almost equal power. The
origin of the soliton fission mechanism is attributed to TPA as
demonstrated in c-Si waveguide\,\cite{Leo-14}, despite the lower
TPA coefficient encountered in a-Si:H. Without TPA, numerical
simulations reveal that solition fission induced by high-order
dispersion leads to two fundamental solitons with unequal pulse
width and amplitude, and thus to a strongly asymmetric output
spectrum. On the contrary, with the TPA alone the two pulses have
the same amplitude and temporal width, resulting in a symmetric
spectrum around the carrier frequency, as shown
in\,\cite{Silberberg-90} for weak TPA, and in agreement with the
shape of the experimental output spectra. Finally, the small
normal GVD peak around 1.9\,$\mu$m is related to the emission of a
dispersive wave and its position is in excellent agreement with
the experiment.

As for SC generation in photonic crystal fibers\,\cite{Dudley-06},
at higher input peak power (13\,W), it can be noticed that the
spectral broadening is approximately symmetrical in the initial
stage of propagation(see Fig.\,\ref{Fig_3}(b)). At about 0.6\,mm,
the temporal compression is maximum and the simulations show that
the input pulse is compressed down to 16\,fs, revealing the
potential of a-Si:H waveguides for on-chip ultrashort pulse
generation. After the compression stage, the spectrum becomes
asymmetric with the emergence of a large peak in the normal
dispersion regime due to the generation of a strong dispersive
wave, and the development of distinct spectral peaks around the
pump wavelength. These peaks are related in the temporal domain,
with the splitting of the initial pulse into five sub pulses with
different temporal duration and peak power. The soliton number at
the input is $N=6.2$, a value close to the number of ejected
pulses. This is in contrast with the results reported in c-Si at
telecommunication wavelengths where the pulse splitting results in
the generation of a small number, largely lower than the initial
soliton number, of equal fundamental solitons, because of the
large TPA experienced in this latter structure\,\cite{Leo-14b}.

After 1\,cm of propagation, the numerically simulated spectrum
qualitatively agrees with the recorded output spectrum. In the
anomalous dispersion region ($\lambda<1720$\,nm), the edge of the
two spectra near the ZDW agree very well and no evidence of Raman
self-frequency shift is visible. This confirms our assumption that
Raman effect does not play a significant role in the SC
generation. The differences in the spectral features might be
attributed to our model, where we only assume instantaneous
refractive and absorptive nonlinearities. In the normal dispersion
regime, numerical simulations show an efficient transfer of energy
to the dispersive wave up to 25\% of the total output energy. In
our experiment, the spectral peaks are lower in that region,
particularly beyond 1.9\,$\mu$m. This might be attributed to a
linear loss larger than expected in that wavelength range.

\section{Coherence} 

An important characteristic of a supercontinuum is its
shot-to-shot stability and, for some applications such as coherent
spectroscopy or frequency metrology, a stable and coherent SC over
its entire spectral bandwidth is highly desirable. Extensive
studies of SC generation in PCF and conventional fibers have shown
that the sensitivity of the SC to noise is inherently related to
the nature of the spectral broadening processes involved. The
coherence property of SC can be viewed as the result of a
competition between soliton fission and modulation
instability\,\cite{Dudley-06}. The spectral coherence of the SC
thus depends on dispersion and nonlinear properties of the
waveguide but also on the pulse duration and peak power of the
seed pulse. It is thus expected the SC reported in the picosecond
regime in a:Si-H waveguides in\,\cite{Safioui-14} to show poor
coherence and the one displayed in Fig.\,\ref{Fig_1} to be
coherent, though the waveguides are similar in both experiments.
Indeed, the broadening mechanism in the latter case is dominated
by soliton fission with a relatively low input soliton number
($N<10$), making the SC less sensitive to pump laser shot noise.

\begin{figure} 
\centerline{\resizebox{110mm}{!}{\includegraphics{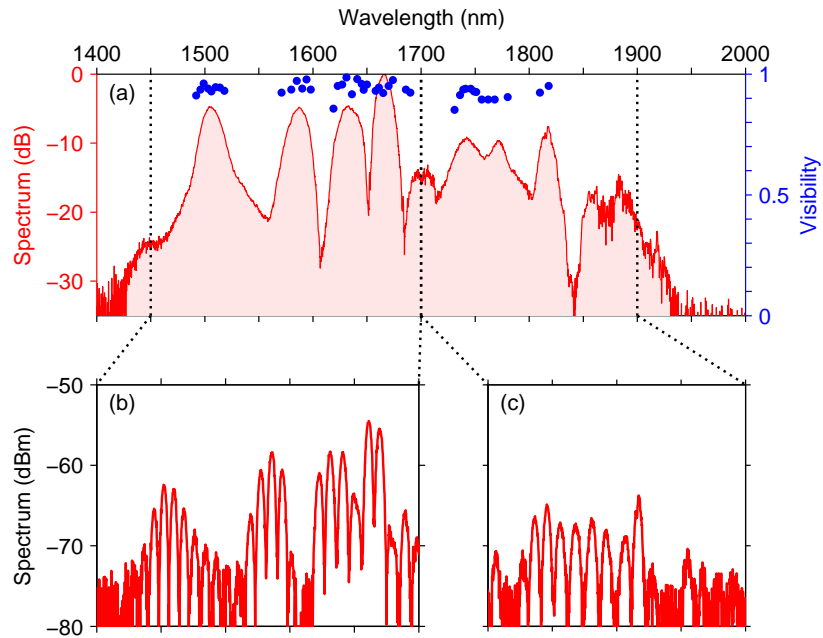}}}
\caption{ Experimental results: (a) Normalized spectrum at the
output of the waveguide. (b) Spectrum at the output of the
interferometer when aligned to maximize the transmission in the
1450-1750 nm spectral region. (c) Same as (b) but for the
1650-1950 nm spectral region. The measured deep fringes indicate
strong phase locking accross the supercontinuum. Note that due to
experimental constraints, the amplitudes in the two arms are not
equal on the whole spectral windows. The fringe visibility thus
gives the minimum value of the first order coherence.}
\label{Fig_4} \vspace{-2mm}
\end{figure}

The coherence properties of the SC generated in the a:Si-H
photonic wires with $<200$\,fs input pulses have been studied
through the degree of first-order coherence
$\vert{\gamma}_{12}^{(1)}(\lambda, t, t+\tau)\vert$, similarly to
our work carried out in c-Si waveguides\,\cite{Leo-14b} . This was
performed by sending the SC collected at the output of the a:Si-H
waveguide into an unbalanced Michelson interferometer where the
delay $\tau$ between the two arms almost matches the pulse delay
between two successive laser pulses\,\cite{Bellini-00}
(see\,\cite{Leo-14b} for further details on the experimental
setup). The spectral interference pattern resulting from two
successive independent SCs was thus measured by a spectrum
analyzer (see insets in Fig.\,\ref{Fig_4}).

The degree of first-order coherence can readily be extracted from
the fringe visibility $V=(\Imax -\Imin)/(\Imax+\Imin)$ where
$I_\mathrm{max, min}$ are the maximum and minimum values of the
fringe pattern: $V=2\vert{\gamma}_{12}^{(1)}\vert[I_1
I_2]^{1/2}/(I_1 + I_2))$ where $I_{1, 2}$ are the intensities in
each arm. The experiment was performed for a SC seeded at a
wavelength of 1620\,nm. This wavelength, closer to the
zero-dispersion wavelength (ZDW), allows to generate an almost
gap-free SC from 1450\,nm to 1900\,nm, together with strong
signals in the normal dispersion regime. Several interference
patterns were recorded, each slightly different due to a small
drift of the fringes between two scans. The visibility at a given
wavelength was measured by looking at the maximum and the minimum
values of the fringes at that wavelength among our data set. The
spectral resolution was 1\,nm, to avoid degradation of the
visibility though ensuring a sufficiently high signal-to-noise
ratio, given the weakness of the signal collected at the output of
the interferometer.

As can be seen in Fig.\,\ref{Fig_4}, the visibility is better than
0.9 almost everywhere in the SC spectrum. The measured values
constitute a lower limit for $\vert{\gamma}_{12}^{(1)}\vert$
because of the intensity difference between the two arms in some
portions of the spectrum, and the noise floor of the OSA which
prevents us to properly measure $\Imin$. This high visibility
shows that the generated SC is highly coherent and stable across
its whole spectral bandwidth as for SC in c-Si\,\cite{Leo-14b}.
This contrasts with the coherence measured in PCF with input
pulses in the range 100-200\,fs in the anomalous dispersion
region\,\cite{Lu-04}.

\section{Material stability}

\begin{figure}[] 
\vspace{-0mm}
\centerline{\resizebox{90mm}{!}{\includegraphics{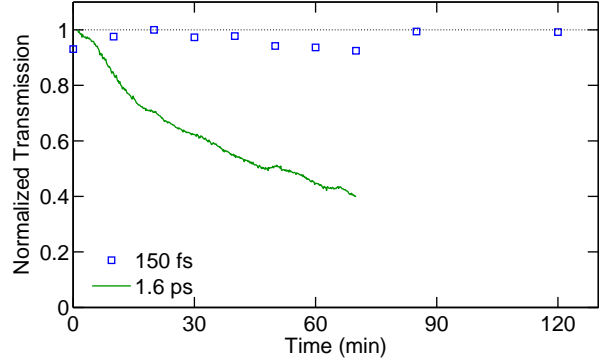}}}
\caption{Transmission, normalized to its maximum value, for
188\,fs and 1.5\,ps input pulse and similar peak power of about
12\,W. The two measurements where taken with the same
post-annealed waveguide.} \label{Fig_5} \vspace{-2mm}
\end{figure}

Hydrogenated amorphous silicon is a commonly used material for
large-area electronics and optoelectronics. However this material
suffers from degradation of its optoelectronic properties under
illumination, known as the Staebler-Wronski effect
(SWE)\,\cite{Staebler-77}. This effect is associated with
light-induced creation of dangling bonds originating from breaking
of weak Si-Si bonds adjacent to a Si-H bond under
illumination\,\cite{Stutzmann-85, Fritzsche-01, Morigaki-07}.
Previous experiments on similar waveguides as in this
work\,\cite{Safioui-14, Kuyken-11}, have reported a degradation of
their nonlinear properties under illumination at telecommunication
wavelengths, presumably from two-photon absorption induced SWE.
However, unlike c-Si, the optical properties of a a:Si-H sample
strongly depend on the fabrication process, which impacts the
extend of disorder in the amorphous structure and the extend of
weak bonds presence, explaining why stable a-Si:H waveguides have
also been reported under similar illumination
conditions\,\cite{Grillet-12}. Moreover, in\,\cite{Safioui-14}, it
has been pointed out that the degradation rate highly depends on
the propagation conditions. Wide waveguides (800\,nm-wide) might
not show any degradation under illumination while smaller
(500\,nm-wide) waveguides on the same chip degrade significantly
after only a few minutes, with picosecond input pulses (see in
Fig.\,\ref{Fig_5}). Interestingly, the spectra of the generated SC
with shorter ($<200$\,fs) input pulses in the same 500\,nm-wide
waveguides show no evidence of any degradation. This is confirmed
in the power transmission recorded under 2\,hours of continuous
illumination as seen in Fig.\,\ref{Fig_5}.

\section{SC generation in a:Si-H and c-Si waveguides}

Few years ago, it has been realized that amorphous silicon shows
better performance regarding instantaneous third order
nonlinearities at telecommunication wavelengths than its
crystalline counterpart\,\cite{Kuyken-11, Tsang-08, Narayanan-10b,
Mehta-10}. Thanks to a larger energy bandgap around 1.6-1.7\,eV,
a-Si:H is characterized by a lower nonlinear absorption
coefficient around 1550\,nm while keeping a large instantaneous
Kerr effect, giving rise to an overall higher nonlinear
figure-of-merit (FOM). By exploiting this high FOM, key
experiments such as wavelength conversion\,\cite{Wang-12,
Mehta-12,  Suda-12}, parametric amplification\cite{Kuyken-11},
supercontinuum generation with picosecond
pulses\,\cite{Safioui-14}, or ultrafast all-optical
modulation\,\cite{Vukovic-13} or waveform
sampling\,\cite{Kuyken-11b}, have been conducted in a-Si:H
structures, and demonstrate its potential for low power ultrafast
silicon based photonic technologies.

\begin{figure}[] 
\vspace{-0mm}
\centerline{\resizebox{110mm}{!}{\includegraphics{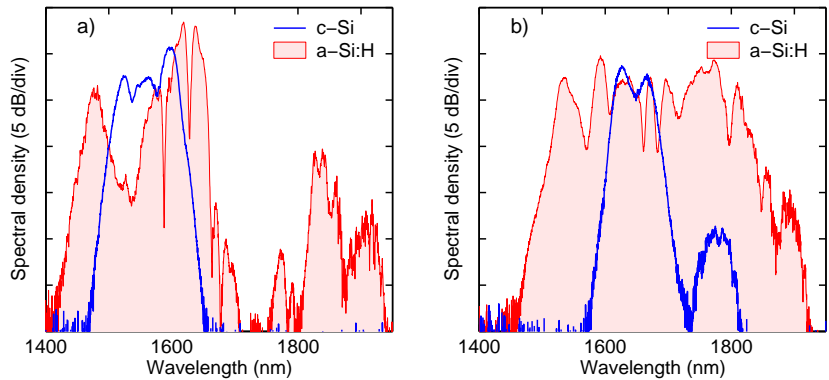}}}
\caption{Supercontinuum generation in similar waveguides made up
of a-Si:H or c-Si. The dispersion curve of these waveguides are
displayed in Fig.\,\ref{Fig_1}. The pump wavelength is
$\lambda_s=1575$\,nm (a) and $\lambda_s=1650$\,nm (b).}
\label{Fig_6} \vspace{-2mm}
\end{figure}

In order to prove the advantage of a-Si:H over c-Si for
supercontinuum generation with femtosecond pulses, we have made a
direct comparison between SCG in both structures. The dispersion
properties of c-Si waveguides, fabricated with the same mask as
for a-Si:H, have been measured with the same setup as discussed in
section\,\ref{Section2}. As can be seen in Fig.\,\ref{Fig_2}, the
500\,nm-wide c-Si and a:Si-H waveguides show similar dispersion
properties and both have their ZDW around 1700\,nm. The
free-carriers effects being negligible in the dynamics of SCG in
our c-Si waveguides\,\cite{Leo-14}, the only fundamental
differences between the two waveguides lie in the real and
imaginary parts of the nonlinear  parameter $\gamma$ in
Eq.\,\ref{Eq1}. Because of large TPA at telecommunication
wavelengths, the SPM induced spectral broadening of the pump pulse
is clearly lower for c-Si than for a-Si:H waveguides (see
Fig.\,\ref{Fig_6}). For a 1575\,nm pump, the dispersive wave
wavelength in the normal GVD regime is too far from the pump
wavelength for this energy transfer to occur in c-Si. On the
contrary, in the a-Si:H photonic wire, the peaks around the DW
wavelength at 1900\,nm are well developed. As previously
discussed, we can even expect these peaks to be higher in shorter
a-Si:H waveguides. At a pump wavelength closer to the ZDW, the DW
became visible in the spectrum at the output of the c-Si
waveguide, while for the a-Si:H one, the SC is at least four times
broader and fully connected. Given the recent results in c-Si
waveguides for shorter pump pulses of 80\,fs\,\cite{Ishizawa-14},
based on numerical simulations, we can expect by pumping our
a-Si:H waveguide at 1550\,nm to easily generate an almost flat SC
between the two dispersive waves around 1200\,nm and 2000\,nm.

\section{Conclusions}

We have experimentally studied the generation of supercontinuum in
hydrogenated amorphous silicon photonic wires in the femtosecond
regime at telecommunication wavelengths. Thanks to an engineered
dispersion through the waveguide width, the waveguide dispersion
is anomalous in the C-band. The numerical simulations based on the
experimentally measured dispersion curve shows that the anomalous
dispersion enable a strong temporal compression down to 15\,fs
starting from a 180\,fs pump pulse. Following the compression, the
pulse experiences high-order soliton fission as well as dispersive
wave generation leading to a broad output spectrum spanning from
1420\,nm to 1950\,nm. The spectral broadening in a-Si:H photonic
wire was demonstrated to be significantly larger than in
crystalline waveguides with similar cross section and dispersion
properties. The agreement between simulations and experiments
shows that a simple model taking high-order dispersion,
instantaneous refractive and absorptive nonlinear effects, and
self-steepening is sufficient to capture the overall dynamics of
the SC generation in our a-Si:H waveguide with femtosecond input
pulses. We have demonstrated that the generated supercontinuum is
highly coherent as expected from soliton fission
dynamics\,\cite{Dudley-06}. Moreover, contrary to previous work in
the picosecond regime, the a-Si:H waveguides have not shown any
material degradation, nor in the SC spectrum nor in the power
transmission.

a-Si:H waveguides thus appear as an interesting platform for
applications requiring on chip coherent low power broad
supercontinuum or ultrashort ($<20$\,fs) pulses at
telecommunication wavelengths, where low-power pump pulse lasers
in the range 100-200\,fs are readily available. Moreover, thanks
to low two-photon absorption, a-Si:H photonic wires are ideally
suitable for the demonstration, in silicon-based technology, of
new phenomena related to optical event horizon and/or dispersive
waves\,\cite{Demircan-11, Choudhary-12}.

\section*{Acknowledgment}

This work was supported by the Belgian Science Policy Office
(BELSPO) Interuniversity Attraction Pole (IAP) project
Photonics@be and by the Fonds de la Recherche Fondamentale
Collective, Grant No. 2.4563.12. Part of this work was carried out
in the framework of the FP7-ERC projects MIRACLE and InSpectra. B.
Kuyken acknowledges the special research fund of Ghent University
(BOF), for a post doctoral fellowship.

\end{document}